\documentclass[%
 aip,
 sd,%
 amsmath,amssymb,
 reprint,%
]{revtex4-1}

\usepackage{graphicx}% Include figure files
\usepackage{dcolumn}% Align table columns on decimal point
\usepackage{bm}% bold math
\begin{document}

\preprint{AIP/123-QED}

\title{Enhancement of critical current density and mechanism of vortex pinning in H$^+$-irradiated FeSe single crystal}% Force line breaks with \\

\author{Yue Sun}
\email[Author to whom correspondence should be addressed.]{Electronic mail: sunyue.seu@gmail.com}
\author{Sunseng Pyon}
\author{Tsuyoshi Tamegai}
 \affiliation{
Department of Applied Physics, The University of Tokyo,Tokyo 113-8656 Japan%\\This line break forced with \textbackslash\textbackslash
}
\author{Ryo Kobayashi}
\author{Tatsuya Watashige}
\author{Shigeru Kasahara}
\author{Yuji Matsuda}
 \affiliation{Department of Physics, Kyoto University, Kyoto 606-8502 Japan}
 \author{Takasada Shibauchi}
 \affiliation{Department of Advanced Materials Science, The University of Tokyo, Chiba 277-8561 Japan}
\author{Hisashi Kitamura}
\affiliation{Radiation Measurement Research Section, National Institute of Radiological Sciences, Chiba 263-8555 Japan}

\begin{abstract}
In this report, we comprehensively study the effect of H$^+$ irradiation on the critical current density, $J_c$, and vortex pinning in FeSe single crystal. It is found that the value of $J_c$ for FeSe is enhanced more than twice after 3-MeV H$^+$ irradiation. The scaling analyses of the vortex pinning force based on the Dew-Hughes model reveal that the H$^+$ irradiation successfully introduce point pinning centers into the crystal. We also find that the vortex creep rates are strongly suppressed after irradiation. Detailed analyses of the critical current dependent pinning energy based on the collective creep theory and extend Maley's method show that the H$^+$ irradiation enhances the value of $J_c$ before the flux creep, and also reduces the size of flux bundle, which will further reduce the field dependence of $J_c$ due to vortex motion.
%
%Valid PACS numbers may be entered using the \verb+\pacs{#1}+ command.
\end{abstract}
\maketitle
Iron-based superconductors (IBSs) display some fascinating fundamental properties for applications, such as reasonably high value of superconducting transition temperature, $T_c$, very high critical field, $H_{c2}$, and relatively small anisotropy.\cite{GurevichRPP} Among the IBSs, iron chalcogenides have stimulated great interests since they are possible candidates to break the $T_c$ record ($\sim$ 55 K) in IBSs. Although the initial $T_c$ of FeSe is below 10 K,\cite{HsuFongChiFeSediscovery} it increases up to 14 K with appropriate Te substitution,\cite{SalesPRB} and 37 K under high pressure.\cite{MedvedevNatMat} \cite{BurrardNatMat} Furthermore, the monolayer of FeSe film grown on SrTiO$_3$ even shows a sign of superconductivity over 100 K.\cite{GeNatMatter} For applications, high quality Te-doped FeSe tapes with transport $J_c$ over 10$^6$ A/cm$^2$ under self-field and over 10$^5$ A/cm$^2$ under 30 T at 4.2 K were already fabricated.\cite{SiWeidongNatComm} In addition, its less toxic nature than iron pnictides is also advantage for applications.

For the application point of view, the value of $J_c$ is a key factor, which is determined not only by material's intrinsic properties but also by extrinsic conditions, like the defects. Thus, the introduction of artificial pinning centers either by chemical or physical methods is effective to enhance the value of $J_c$. The chemical method introduces extended defects, like Y$_2$O$_3$ nanoparticles in bulk cuprates.\cite{JinPRBYBCO} The physical method is usually performed by particle irradiations, such as the point defects caused by proton irradiation and columnar defects by heavy-ion irradiation. The physical method is more advantageous to probe the pinning mechanism because it is easy to control the number and type of pinning centers without affecting the structure of the crystal. For IBSs, both methods are proved to be effective to the enhancement of $J_c$.\cite{LeeNatMat,FangNatCom,NakajimaPRBirra,HaberkornPRBBaCo,KihlstromAPL,TamegaiSUST,MasseeSA} However, until now, attempts have been made mostly in iron pnictides, especially in the "122" phase since high-quality single crystals are available. For FeSe, such study is still left unexplored because of the difficulty in growing high-quality single crystals. Actually, the study of irradiation effect in FeSe is not only important to the enhancement of $J_c$ for applications but also crucial for the understanding of pinning mechanism since FeSe possesses some unique characteristics. It has the simplest structure, composed of only Fe-Se layers, and is also a clean system free from doping introduced inhomogeneities and charged quasi-particle scattering because of its innate superconductivity.\cite{HsuFongChiFeSediscovery}

Recently, high-quality and sizable single crystals of FeSe have been grown.\cite{BöhmerPRB} In this report, we present the study of H$^+$ irradiation effect on FeSe single crystal. Introduction of defects into FeSe using 3-MeV H$^+$ results in the enhancement of $J_c$ by a factor of more than two, which is explained by the successful introduce of point pinning centers into the crystal. Vortex dynamics study reveals that the proton irradiation enhances the value of critical current density before the flux creep, and also reduces the size of flux bundle, which will further suppress the strong field dependence of $J_c$ from vortex motion.

High quality single crystals of tetragonal $\beta$-FeSe were grown by the vapor transport method as described elsewhere.\cite{KasaharaPNAS} Our previous report has shown the high quality of the grown FeSe single crystal, which exhibits $T_c$ $\sim$ 9 K with the residual resistivity ratios RRR $>$ 40. Besides, scanning tunneling microscope (STM) observations also manifested that the crystal contains extremely small level of impurities/defects.\cite{KasaharaPNAS,Watashigearxiv} Those results ensure that our irradiation experiment was performed on a clean crystal with less influence from second phase or inhomogeneities. FeSe crystals were cleaved to thin plates with thickness $\sim$ 25 $\mu$m along the $c$-axis, which is much smaller than the projected range of 3-MeV H$^+$ for FeSe of $\sim$ 50 $\mu$m, calculated by the stopping and range of ions in matter-2008.\cite{irradiationrange} To avoid the possible sample-dependent influence, all the measurements were performed on one identical piece of crystal, which was divided into two parts; pristine and irradiated samples. The 3-MeV H$^+$ irradiation was performed parallel to the $c$-axis at National Institute of Radiological Science-Heavy Ion Medical Accelerator in Chiba with a total dose of 5 $\times$ 10$^{16}$ /cm$^2$. Magnetization measurements were performed by using a commercial SQUID magnetometer. After the irradiation, the value of $T_c$ is almostly unchanged, which is similar to the case of Ba$_{1-x}$K$_x$Fe$_2$As$_2$.\cite{KihlstromAPL}

\begin{figure}\center
\includegraphics[width=8.5cm]{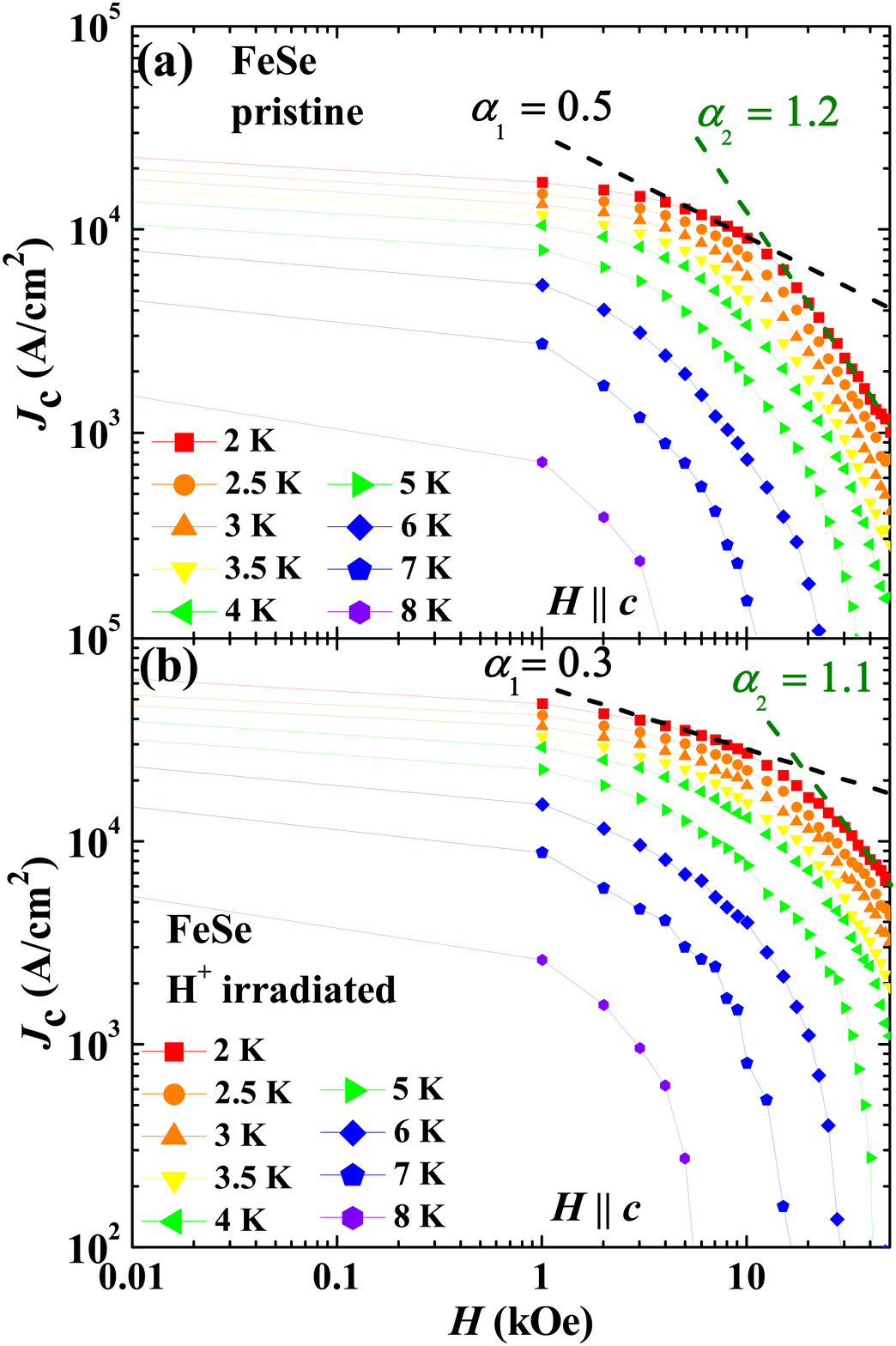}\\
\caption{(Color online) Magnetic field dependence of critical current densities with $H$ $\|$ $c$ for (a) the pristine and (b) H$^+$-irradiated FeSe. The dashed lines show the power-law decay of $H^{-\alpha}$.}\label{}
\end{figure}

Fig. 1 shows the magnetic field dependence of $J_c$ for the (a) pristine and (b) H$^+$-irradiated FeSe single crystal obtained by using the extended Bean model:\cite{Beanmodel}
\begin{equation}
\label{eq.1}
J_c=20\frac{\Delta M}{a(1-a/3b)},
\end{equation}
where $\Delta$\emph{M} is \emph{M}$_{down}$ - \emph{M}$_{up}$, \emph{M}$_{up}$ [emu/cm$^3$] and \emph{M}$_{down}$ [emu/cm$^3$] are the magnetization when sweeping the field up and down, respectively, \emph{a} [cm] and \emph{b} [cm] are sample widths (\emph{a} $<$ \emph{b}). It is obviously that H$^+$ irradiation enhances the value of self-field $J_c$ at 2 K from $\sim$ 3 $\times$ 10$^{4}$ to $\sim$ 8 $\times$ 10$^{4}$ A/cm$^2$.

$J_c$ changes little below 1 kOe in the pristine sample, which is followed by a power-law decay $H^{-\alpha}$ in the field range of 4 - 10 kOe with $\alpha_1$ $\sim$ 0.5. Such a power-law dependence of $J_c$ is also observed in most of IBSs, which is attributed to strong pinning by sparse nm-sized defects as in the case of YBCO films.\cite{vanderBeekPRB2002} Such a result is consistent with the STM observation, where randomly distributed defects with their effect range in a few nm scale are dispersed.\cite{KasaharaPNAS} After that, the decaying rate of $J_c$ increases to $\alpha_2$ $\sim$ 1.2. Such behaviors may be explained by the small amount of strong pinning centers with density less than one per 2000 Fe atoms as observed by STM.\cite{KasaharaPNAS} In this case, all the pinning centers will be easily occupied by the flux above some characteristic field. Above 10 kOe, the pinning force $F_p$ will keep constant in spite of the increase in $H$. Thus, the value of  $J_c$ will decrease with the rate of $H^{-1}$ since $F_p$ = $\mu_0H$$\cdot$$J_c$. After H$^+$ irradiation, $J_c$ also shows field insensitive behavior at small field and $H^{-1}$ decaying behavior at fields larger than 10 kOe, similar to the pristine sample. However, in the field range of 4 - 10 kOe, $J_c$ decays with field at a rate of $H^{-0.3}$ rather than the $H^{-0.5}$ behavior observed in the pristine sample. The change of $J_c$ decaying with field from $H^{-0.5}$ to $H^{-0.3}$ was also observed in H$^+$-irradiated Ba(Fe$_{0.93}$Co$_{0.07}$)$_2$As$_2$ and Ba$_{0.6}$K$_{0.4}$Fe$_2$As$_2$.\cite{TaenPRBBaCoirra,TaenSUST}

\begin{figure}\center
\includegraphics[width=8.5cm]{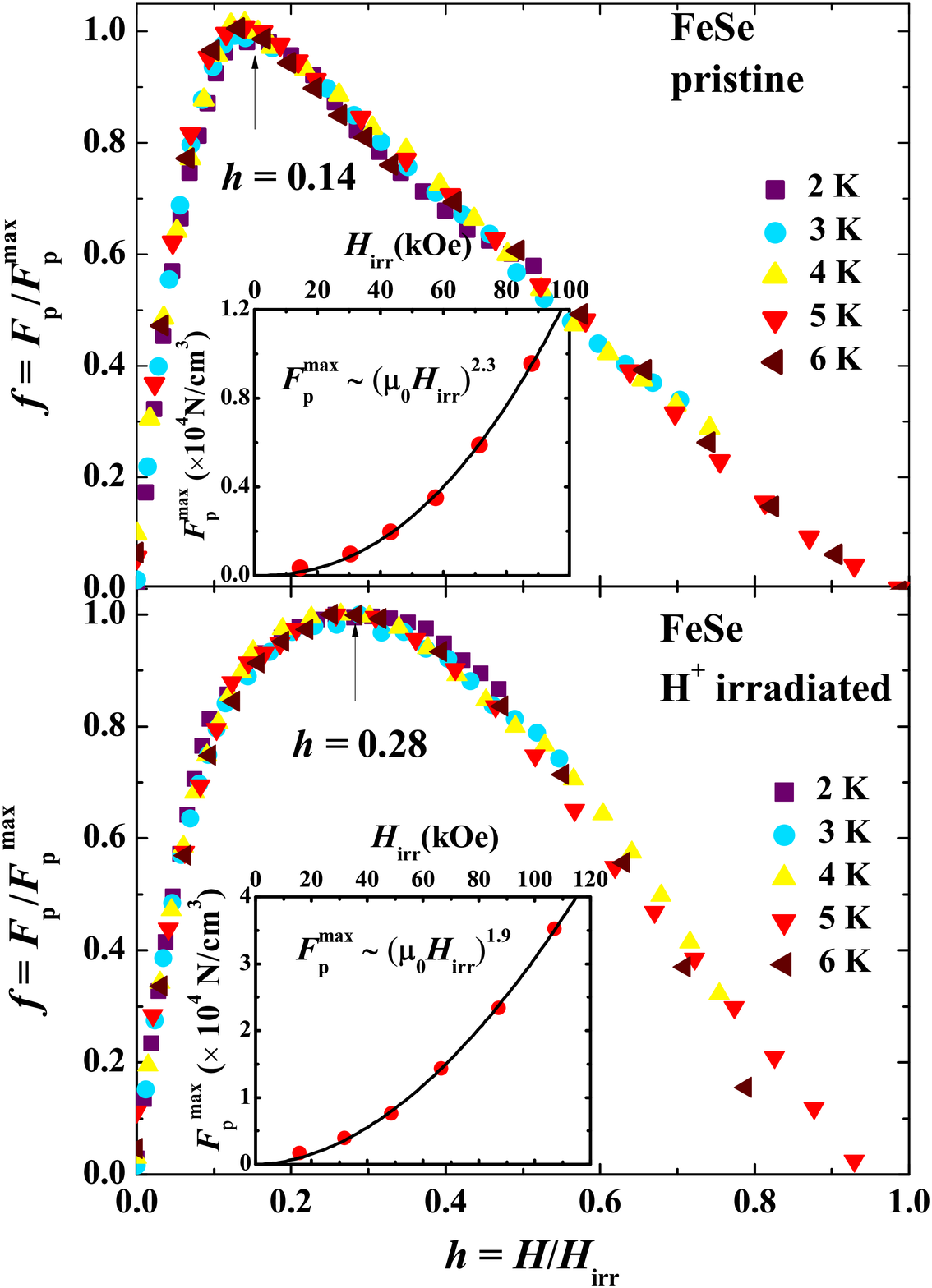}\\
\caption{(Color online) Normalized flux pinning force $f$ = $F_p$/$F_p^{max}$ as a function of the reduced field $h$ = $H$/$H_{irr}$ at different temperatures for (a) pristine and (b) H$^+$-irradiated FeSe. Insets show $F_p^{max}$ as functions of $H_{irr}$.}\label{}
\end{figure}

In order to see the H$^+$-irradiation effect more clearly and gain more insight into the vortex pinning, the normalized vortex pinning forces $f$ = $F_p$/$F_p^{max}$ as a function of the reduced field $h$ = $H$/$H_{irr}$ at different temperatures were shown in Figs. 2(a) and (b) for the pristine and H$^+$-irradiated FeSe, respectively. The pinning force $F_p$ was obtained from critical current density by $F_p$ = $\mu_0H$$\cdot$$J_c$, and $F_p^{max}$ corresponds to the maximum pinning force. $H_{irr}$ is the irreversibility field, which is obtained from the linear extrapolation of $J_c^{1/2}$ - $\mu_0H$ curves to the zero value of $J_c$. It is obvious that $f$ for the pristine and H$^+$-irradiated FeSe all falls into one curve. The peak position of the overlapped curves for the pristine sample locates at the value of $h$ = 0.14. After H$^+$ irradiation, the peak position of $f$ was changed to $h$ = 0.28, which is close to the values of 0.33 for the core normal point pinning  according to the Dew-Huges model.\cite{DewHughes} Moreover, insets of Fig. 2(a) and (b) show the curves of $F_p^{max}$ vs $H_{irr}$ for the pristine and H$^+$-irradiated FeSe. Obviously, the magnitude of $F_p^{max}$ was enhanced, and the scaling parameter $\alpha$ of $F_p^{max}$ $\propto$ $H$$^\alpha$ is around 1.9 for the irradiated crystal, which is close to the theoretical value of 2 for the core normal point-like pinning.\cite{DewHughes} Thus, the peak position change in $f$ indicates that the H$^+$ irradiation successfully introduce point pinning centers into FeSe single crystal, which enhances the pinning force and critical current density.

To get more comprehensive and quantitative understanding of the H$^+$-irradiation effect to the vortex dynamics of FeSe single crystal, we carefully traced the decay of magnetization with time $M(t)$ originated from the flux creep for more than one hour, where $t$ is the time from the moment when the critical state is prepared. The normalized magnetic relaxation rate $S$ is defined by $S$ $\equiv$ $\mid$dln$M$/dln$t$$\mid$. In these measurements, magnetic field was swept more than 5 kOe higher than the target field before starting measurements. Fig. 3(a) shows the temperature dependence of the normalized magnetic relaxation rate $S$ at 500 Oe. (larger than the self-fields of $\sim$ 100 Oe and $\sim$ 300 Oe for the pristine and H$^+$-irradiated crystals, respectively.) $S$ for both crystals shows an obvious temperature insensitive plateau in the intermediate temperature region with a relatively large vortex creep rate. The plateau and large vortex creep rate were also observed in YBa$_2$Cu$_3$O$_{7-\emph{$\delta$}}$, \cite{Yeshurunreview} and other IBSs,\cite{ProzorovPRBBaCoJc} which can be interpreted by the collective creep theory. \cite{Yeshurunreview} On the other hand, after the H$^+$-irradiation, the magnitude of $S$ is suppressed to half of the value of the pristine sample, and the plateau behavior becomes more obvious. Similar suppression of $S$ can be also seen in its field dependence as shown in Fig 3(b), which shows typical results at 2 K.

\begin{figure}\center
\includegraphics[width=8.5cm]{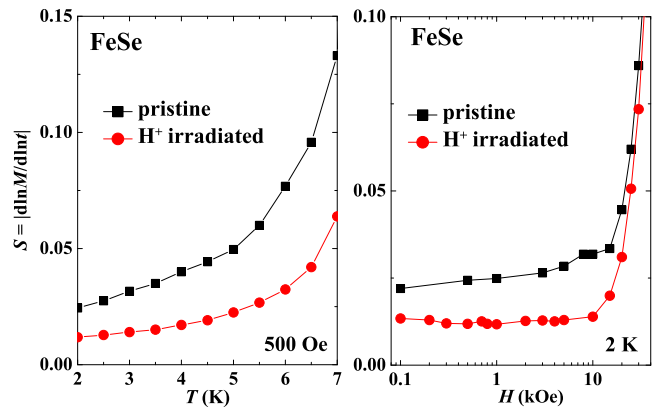}\\
\caption{(Color online) (a) Temperature dependence of the normalized magnetic relaxation rate $S$ for the pristine and H$^+$-irradiated FeSe under 500 Oe. (b) Magnetic field dependence of $S$ for the pristine and H$^+$-irradiated FeSe at 2 K.}\label{}
\end{figure}

Figs. 4(a) and (b) show the effective pinning energies $U^*$ (= $T/S$) as a function of inverse current density 1/$J$ for the pristine and H$^+$-irradiated FeSe, respectively. According to the collective creep theory, the slope $\mu$ for the $U^*$ - 1/$J$ relation in double logarithmic plot contains information about the size of the vortex bundle. In a three-dimensional system, it is predicted as $\mu$ = 1/7, (1) 5/2, 7/9 for single-vortex, (intermediate) small-bundle, and large-bundle regimes, respectively \cite{Blatterreview,FeigelmanPRL}. The evaluated value of $\mu$ for the pristine crystal is $\sim$ 0.71 as expected for collective creep by large bundles. Contrary to the above prediction of $\mu$ $>$ 0, a negative slope with value $\sim$ -0.81 is obtained at small $J$. The negative slope is often denoted as $p$ in plastic creep theory, which is thought to lead to faster escape of vortices from the superconductors.\cite{AbulafiaPRL} The crossover is persistent after H$^+$ irradiation. However, the value of $\mu$ increases to 1.0, which indicates that the vortex creep in the irradiated crystal is in the form of  intermediate bundle.

\begin{figure}\center
\includegraphics[width=8.5cm]{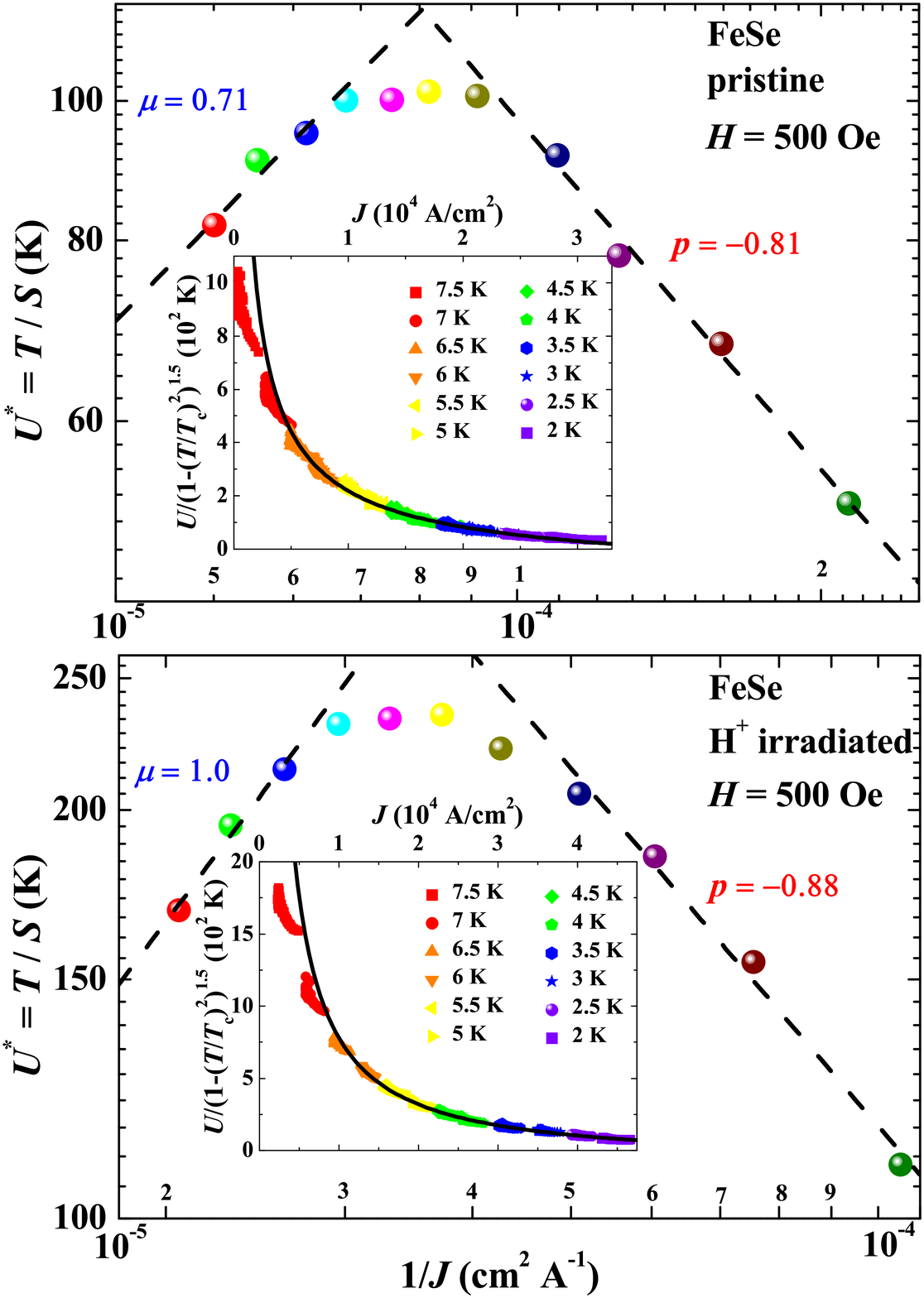}\\
\caption{(Color online) Inverse current density dependence of effective pinning energy \emph{U}$^*$ at 500 Oe for (a) pristine and (b)  H$^+$-irradiated FeSe single crystal. Insets show the current density dependence of flux activation energy \emph{U} constructed by the extended Maley's method.}\label{}
\end{figure}

To get more quantitative insight of the variation in vortex pinning, we analyze the $U$ - $J$ relation by the extended Maley's method.\cite{MiuMayle} We find that all the curves can be well scaled together as shown in the insets of Figs. 4(a) and (b) for the pristine and H$^+$-irradiated FeSe, respectively. The solid lines indicate the power-law fitting by\cite{FeigelmanPRB}
\begin{equation}
\label{eq.3}
U(J)=\frac{U_0}{\mu}[(J_{c0}/J)^\mu-1]
\end{equation}
to the large $J$ region, where  $U_0$ and $J_{c0}$ is the flux activation energy and critical current density in the absence of flux creep, respectively. Deviation of the data from the fitting line in the small $J$ region is reasonable since vortex creep is plastic there. The fitting gives $\mu$ = 0.72,  $U_{0}$ = 91.1 K, $J_{c0}$ = 4.0 $\times$ 10$^4$ A/cm$^2$ for the pristine crystal, and $\mu$ = 1.09, $U_{0}$ = 96.2 K, $J_{c0}$ = 8.1 $\times$ 10$^4$ A/cm$^2$ for the irradiated one. The values of $\mu$ obtained from the extended Maley's method are very close to those evacuated in the main panel of Fig. 4 for both crystals, which confirms the correctness of the present analyses. The observed changes in $J_{c0}$ and $\mu$ show that the H$^+$ irradiation enhances the critical current density without flux creep, and also reduce the size of flux bundle to suppress the reduction of current density from vortex motion.

In conclusion, we report that 3-MeV H$^+$ irradiation can enhance the $J_{c}$ of FeSe single crystal more than twice by introducing extra point pinnings. Magnetic relaxation measurements show that the vortex creep rate is strongly suppressed after the irradiation. Detailed analyses of the critical current dependent pinning energy based on the collective creep theory and extend Maley's method demonstrate that the H$^+$ irradiation enhances the value of $J_{c}$ before the flux creep starts, and also reduces the size of flux bundle from large to intermediate. The reduction of the size of flux bundle will further suppress the field dependence of $J_c$ due to vortex motion.

Y.S. gratefully appreciates the support from JSPS.

% Produces the bibliography via BibTeX.
\bibliography{aipsamp}

\end{document}